\documentclass[lettersize,journal]{IEEEtran}
\usepackage{amsmath,amsfonts}
\usepackage{algorithmic}
\usepackage{algorithm}
\usepackage{array}
\usepackage[caption=false,font=normalsize,labelfont=sf,textfont=sf]{subfig}
\usepackage{textcomp}
\usepackage{stfloats}
\usepackage{url}
\usepackage{verbatim}
\usepackage{graphicx}
\usepackage{cite}
\usepackage{comment}
\usepackage[center]{caption}

\begin{document}

\title{Decoding Motor Imagery \\with various Machine Learning techniques}

\author{Giovanni Jana, Corey Karnei, Shuvam Keshari}

\maketitle

\begin{abstract} 
Motor imagery (MI) is a well documented technique used by subjects in BCI (Brain Computer Interface) experiments to modulate brain activity within the motor cortex and surrounding areas of the brain. In our term project we conducted an experiment in which the subjects were instructed to perform motor imagery that would be divided into two classes (Right and Left). Experiments were conducted with two different types of electrodes (Gel and POLiTag) and data for individual subjects was collected. In this paper we will apply different machine learning (ML) methods to create a decoder based on offline training data that uses evidence accumulation to predict a subjects intent from their modulated brain signals in real time.   
\end{abstract}

When working with BCIs, there is a direct correlation between how your decoder performs and how well your system is able to provide feedback to the user. This is motivation for researchers to continuously explore new methods of decoding brain signals for use in BCI application. Despite an advancement in technology, there are several key features of EEG signals that make them difficult to decode accurately. One feature is that they are non-stationary which means that from one session to another they can present differently. Because of this when decoding, channels must be chosen that display stability over time. Another feature is that EEG signals can present differently for different subjects. This means that a decoder that is built and trained for a certain individual will not perform well on a different individual. Due to this, we will be building decoders for each subject as well as testing multiple ML techniques for each subject. This study has an additional issue; namely the electrodes being used. Non-invasive EEG recordings often utilize an electrolyte gel in order to minimize impedance between an electrode and a subject's scalp. This was possible for our Gel experiments, however using a new type of electrodes resulted in increased impedance per channel that was dependent on the oiliness of a subjects scalp.
\section{Introduction}

\begin{figure}[!h]
\centering
\includegraphics[width=3.4in]{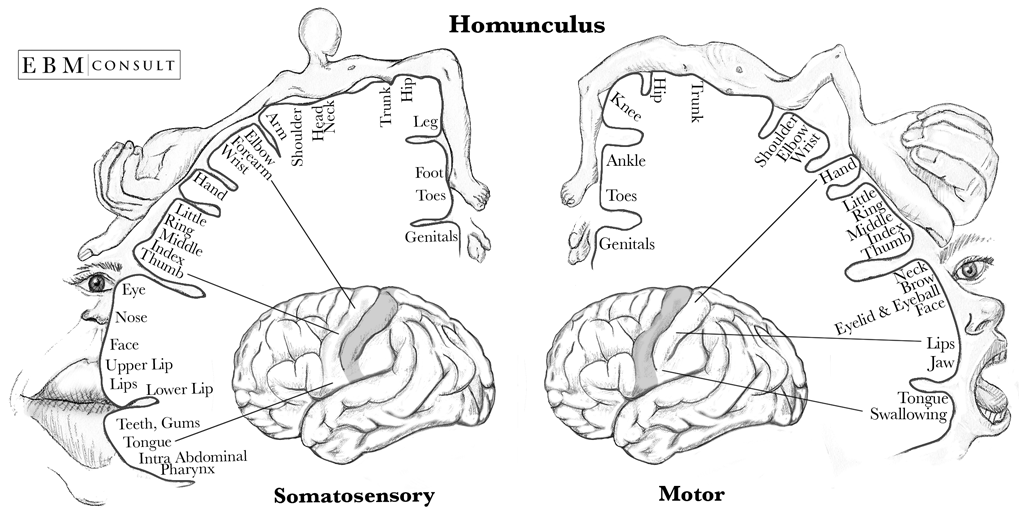}
\caption{Homunculus model}
\label{homonculus}
\end{figure}

Despite all of these obstacles, there are some assumptions we can make when designing our experiment. Literature has shown that there are some common characteristics of EEGs produced by subjects performing MI tasks. Roth et al. \cite{ref1} showed that there is evidence of brain activity in the primary cortex during motor imagery tasks. Another commonality is that the sensory motor rhythms generated during MI occur within the alpha (8-12Hz) and beta (12-30Hz) bands as shown by McFarland et al. \cite{ref2}. Specifically for our experiment we will be conducting motor imagery tasks that involve the arms and hands. The homunculus model in Figure \ref{homonculus}. depicts a representation of what areas of the somatosensory and motor cortex correspond to the hands and arms. Using this knowledge, electrodes will be placed in positions that can capture these areas. For BCI systems that involve the motor imagery task, this has been extensively studied which allows researchers to use a smaller subset of the total channels while still capturing the data necessary for classification.

\section{Experimental Setup}
\begin{figure}[!h]
\centering
\includegraphics[width=3.4in]{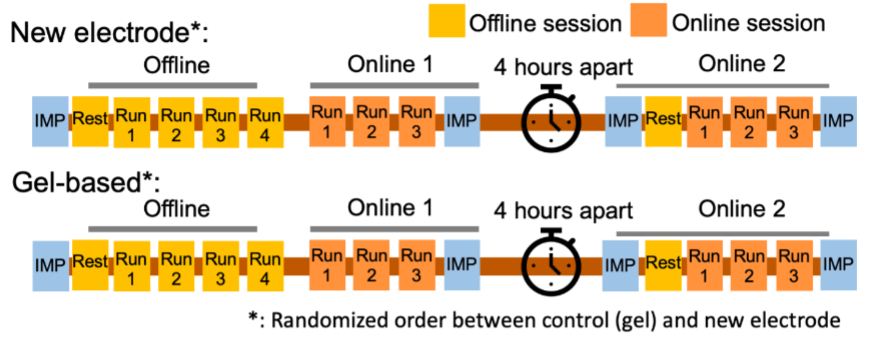}
\caption{Recording Session Structure}
\label{fig_1}
\end{figure}

The experiment was conducted over the course of two weeks. During this time three subjects each participated in recording sessions using both Gel-based and new POLiTag experimental electrodes. The structure of a recording can be seen in Figure \ref{fig_1}. 

\begin{figure}[!h]
\centering
\includegraphics[width=3.4in]{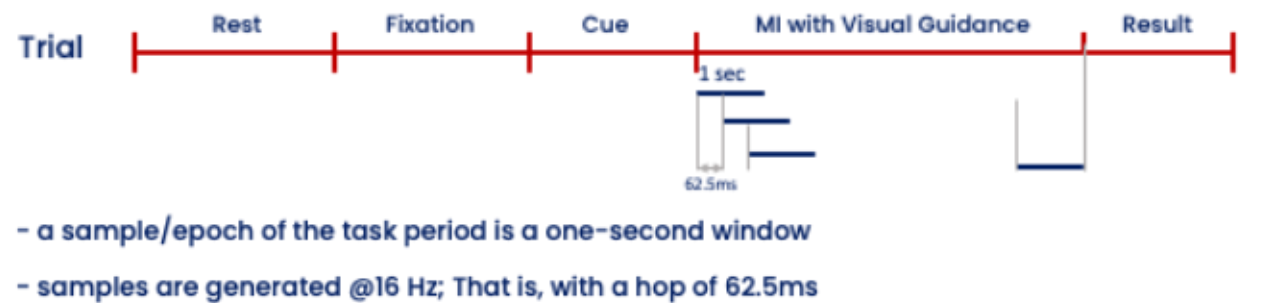}
\caption{Single Trial Structure}
\label{fig_2}
\end{figure}

On a given day, for one type of electrodes there was a single offline session as well as two online sections. Before the first offline session there was an impedance check conducted in order to ensure that all 13 channels were operating within tolerable impedance ranges. This varied between the types of electrodes because for POLiTag impedance could be heavily influenced by whether a subject had an oily or dry scalp, whereas the Gel electrodes used a conductive gel layer to ensure low impedance values.  
\subsection{Offline}
The first session that was run on a given day was the offline session. In this session, resting brain activity was recorded before runs were started. After this the subject would do four runs. A run in this case is 20 trials, where a trial is structured as seen in Figure \ref{fig_2}. These trials are divided equally into 10 trials for the left and right cues. After a period of rest and a fixation cross, the subject would be presented with a cue to perform motor imagery of either a right or left hand task. The subjects were suggested that they could imagine a bicep curl movement for the right hand side and a lateral raise movement for the left hand side. There would be a bar on the screen that shows feedback based on decoder performance, however for the offline session this feedback bar was simulated. After all 20 trials were conducted, that would mark the end of a single run. This was repeated four times in order to form a data set to build the decoder with.

\subsection{Building a Decoder} 
After the offline sessions, a decoder was built using data from the four offline runs. This was done by running a pre-made script and selecting relevant channels through visual inspection. MI is typically seen in modulation between the frequencies of 4-30Hz. Using this knowledge 20 features were selected that were each a channel and a frequency within the aforementioned range. These selections would be used in order to generate the decoder utilized in the following online runs.

\subsection{Online 1}
The first online session was structured similar to the offline section, however only three runs were conducted. Each run was structured the same as before, however now the section of MI with visual guidance (Figure \ref{fig_2}.) was displaying results based on the generated decoder. Using evidence accumulation, the bar would incrementally change towards the side that the decoder was detecting based on the subjects brain activity modulation. Once a certain threshold was reached (or the system timed out) a result would be presented. The result would be either a check mark or x based on whether the subject met the threshold for the correct side of the cue given. It is important to note that this threshold could be adjusted between runs in order to reduce frustration. At the end of this session, another impedance measurement would be conducted to see if there was significant change over the course of the two morning sessions. 

\subsection{Online 2}
At least four hours after the first online session, the second online session was conducted. Because brain signals are non stationary it is important to get data at a different time to help account for this fact. One big difference with online session 2 is that for the Gel the electrodes and gel had to be re-applied between the online sessions, while the POLiTag stayed on for all four hours. The structure of the session began with an impedance check and rest measurement. The three runs were conducted in the same way as the first online session, and following them one last impedance check was conducted. 

\section{Results}
\subsection{Data Collection and Pre-Processing}
\begin{figure}[!h]

\centering
\includegraphics[width=3.4in]{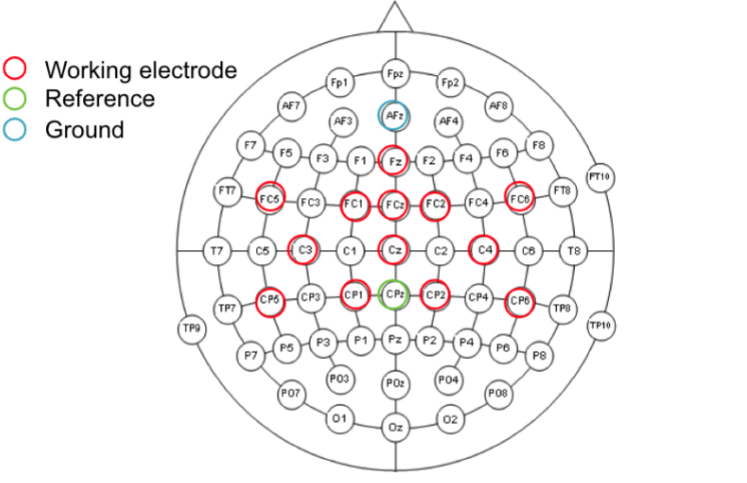}
\caption{Electrode Configuration}
\label{fig_3}
\end{figure}

During the experiments, EEG data was collected continuously over the course of each run. Thirteen electrodes were used placed at locations in Figure \ref{fig_3}.  based on the 10-20 international system. Each electrode was being sampled at a frequency of 512Hz. Offline trials had a consistent length, however online trials were inconsistent because of the time it could take to reach a threshold or timeout. Before any data analysis was done, all data was temporally filtered using a Butterworth filter from 4-30Hz and spatially filtered using Common Average Referencing (CAR) to improve the signal to noise ratio.

Along with the EEG data, it was also important to capture the time at which certain events occurred during a given run. There were different labels for events, however the one we are concerned with for training a machine learning model is the label for when continuous feedback starts. Using this information along with the filtered signals we were able to extract only the portion of signal that occurred during this event for all trials.

A series of one second windows were then extracted from each trial, with a 62.5ms step between windows. These windows each represent one 'sample' in time and are the smallest unit of data for which the machine learning models make predictions. 

\subsection{Feature Selection}

At a sampling rate of 512Hz, each one second window contains 512 samples with 13 electrode channels recorded at each sample. This gives the data over 6,000 total features. This highly dimensional data is likely to be both slow and unstable if applied naively in the context of machine learning models. For this reason we explored multiple techniques to reduce the dimensionality of our data.

\subsubsection{Principal Component Analysis}

Principal component analysis (PCA) is a popular technique for analyzing datasets containing a high number of dimensions. It is a statistical technique for reducing the dimensionality of a dataset by linearly transforming the data into a new coordinate system where (most of) the variation in the data can be represented with fewer dimensions than the initial data.

\begin{figure}[!h]
\centering
\includegraphics[width=3.4in]{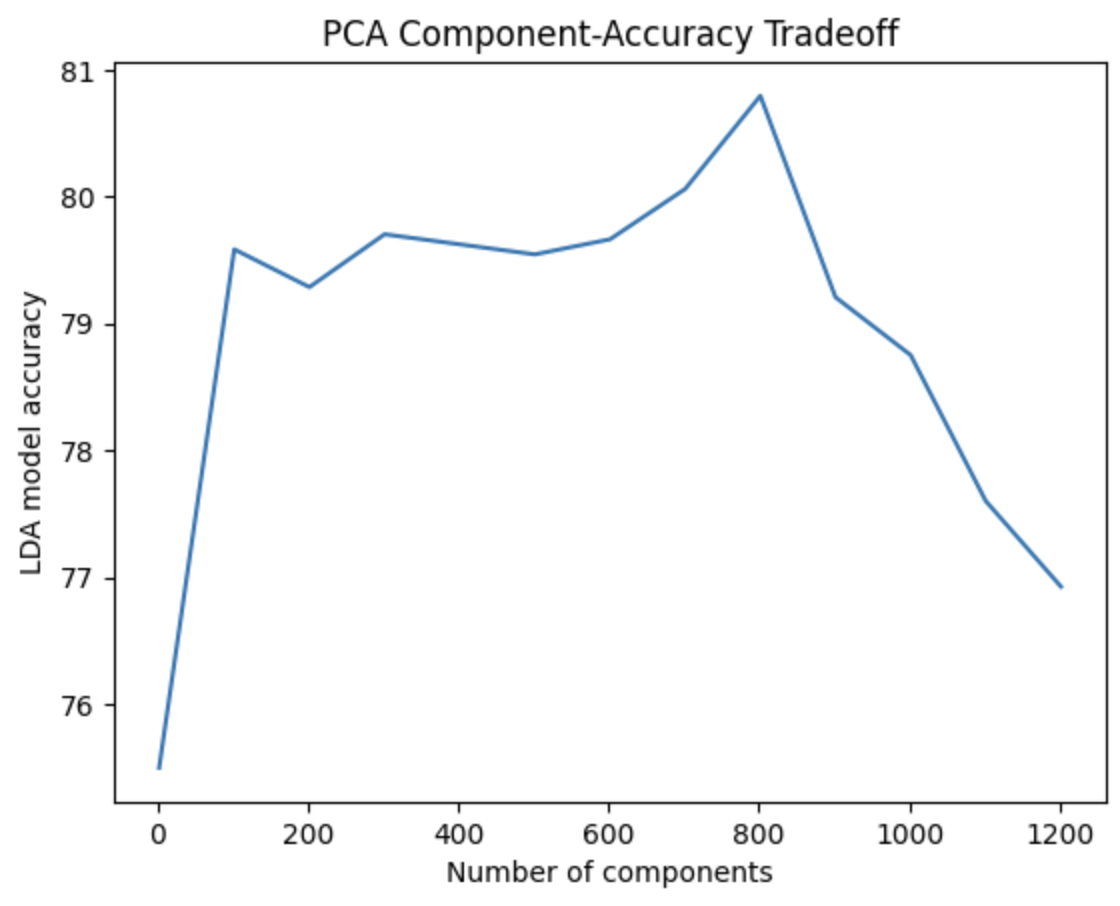}
\caption{Trade-off between number of components remaining after PCA and the accuracy of a linear discriminant analysis model.}
\label{fig_5}
\end{figure}

\begin{figure}[!h]
\centering
\includegraphics[width=3.4in]{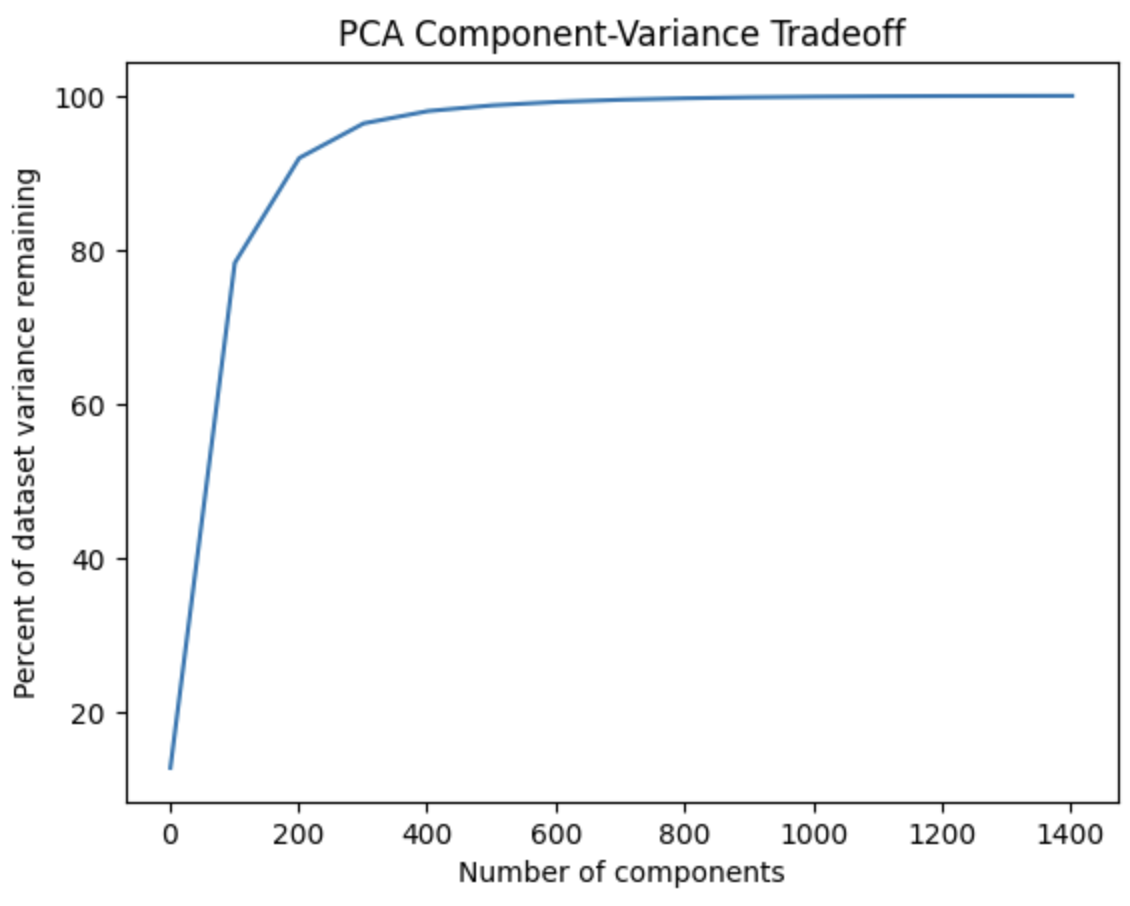}
\caption{Trade-off between number of components remaining after PCA and the percent of variance retained from the original dataset.}
\label{fig_4}
\end{figure}

Figure \ref{fig_4} shows how much variance from the original dataset is retained as the number of remaining dimensions is reduced. To decide which number of remaining components to use in the machine learning pipeline, we used run-wise cross validation of the data at various values of principal components. Figure \ref{fig_5} shows the results of this using a linear discriminant analysis machine learning model. Ultimately the number of remaining components was chosen to be 800. These results are derived using subject 2's Gel offline session, each other subject had an identical analysis done to determine the best PCA values for each subject. The Gel sessions had their best PCA values between 500-800 depending on the subject while the POLiTag sessions had their best values between 100-200 remaining features.

An experiment was tried where the number of components would be iteratively reduced, from more than 6000 to 4000, then from 4000 to 1500, and finally from 1500 to 800. This was ultimately fruitless as our results seemed to indicate that there was no benefit to this compared to reducing directly to the desired value. In fact there were instances where this proved less effective than reducing directly to the desired value.

\subsubsection{Power Spectral Density}

Power spectral density (PSD) is the measure of signal's power content versus frequency. We applied Welch's method to estimate the the spectral density, reducing the dimensionality of each window down to 129. We investigated using PCA to further reduce the dimensionality of the PSD features, but the experiments indicated that there this provided little to no accuracy gain. 

Both of these dimensionality reduction methods are contrasted in the following section about model selection to see which one provides the best performance.

\subsection{Model Selection}

We experimented with a number of different machine learning models when trying to decide which one should be used in our final decoder. We tried a number of prepackaged linear models included in the sklearn package along with some attempts to get a convolutional neural network running. These sets of experiments are described below.

\subsubsection{Convolutional Neural Network} 
The data had an original dimensionality of 512 samples x 13 channels. For the four models just described, this shape was flattened to a 1-D vector of size 6656. However, we wondered if by maintaining this original rectangular shape and using a convolutional neural network, we could get comparable or better accuracy to that of the linear models. 

\begin{figure}[!h]
\centering
\includegraphics[width=3.4in]{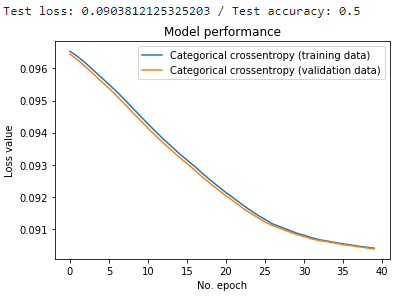}
\caption{Loss function on the training and validation sets}
\label{fig_7}
\end{figure}

We tried different depth of layers for the network, starting with 3 layers. Finally, we created a network with 6 convolutional layers (each accompanied with batch normalization layers and relu activation layers). This network was trained on the data from the offline session that had a total of 3780 'time windows' and their corresponding predictions. The test set consisted of 1260 'time windows'. A total of 40 epochs gave the following results as shown in Figure \ref{fig_7}. As can be seen the results are not very different from random chance. A similar observation was made for networks of different depths as well.

One idea to change the way we were training the network was modifying the training data. Instead of feeding numbers to the model (derived from the 'time window' of the EEG signals), we can create plots of the time windows as shown in Figure \ref{fig_9}, and then feed those images of the plots to the Convolutional neural network. Hence for the above network, we would be training it on 3780 images. It then becomes an image classification problem, for which we have deployable pre-trained networks already available. One limitation of this approach could be the imbalance in classes in the training set that could lead to biases in the network. Another aspect to consider here is that since there are 13 channels, so we would get 13 images for the plots of the 'time windows'. We could feed all 13 channels to the network as a 'superimage' (as opposed to the 3 RGB channels ususally dealth with in colored images), or we could generate topoplots of the time windows and input that stack of topoplot channels as well. Either of these methods could be explored in future, as these were prohibitively time intensive.

To address the question of how to generate the images in real time, we thought of using plots in real time. As soon as we get the data for the 1 second window, we generate an image of that and feed it to the convolutional neural network. It is worth noting that while training the entire model is time intensive, inferring predictions from a trained model is not.

\begin{figure}[!h]
\centering
\includegraphics[width=3.4in]{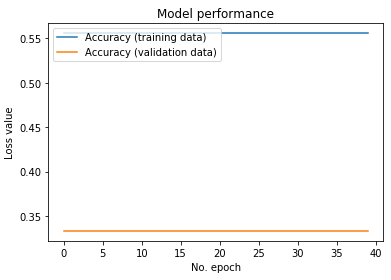}
\caption{Model performance: Accuracy}
\label{fig_8}
\end{figure}

\begin{figure}[!h]
\centering
\includegraphics[width=3.4in]{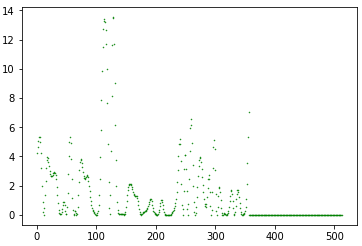}
\caption{An example of the plot of 1 'time window' from the training set}
\label{fig_9}
\end{figure}

\subsubsection{Scikit Learn Models}

We selected collection of 4 models and contrasted them across the different subjects and different feature selection techniques. For this step, no fine-tuning of model parameters was done. The models chosen were random forest, linear discriminant analysis, support vector machine, and multi-layer perception. 

\begin{figure*}[!h]
\centering
\includegraphics[width=6.8in]{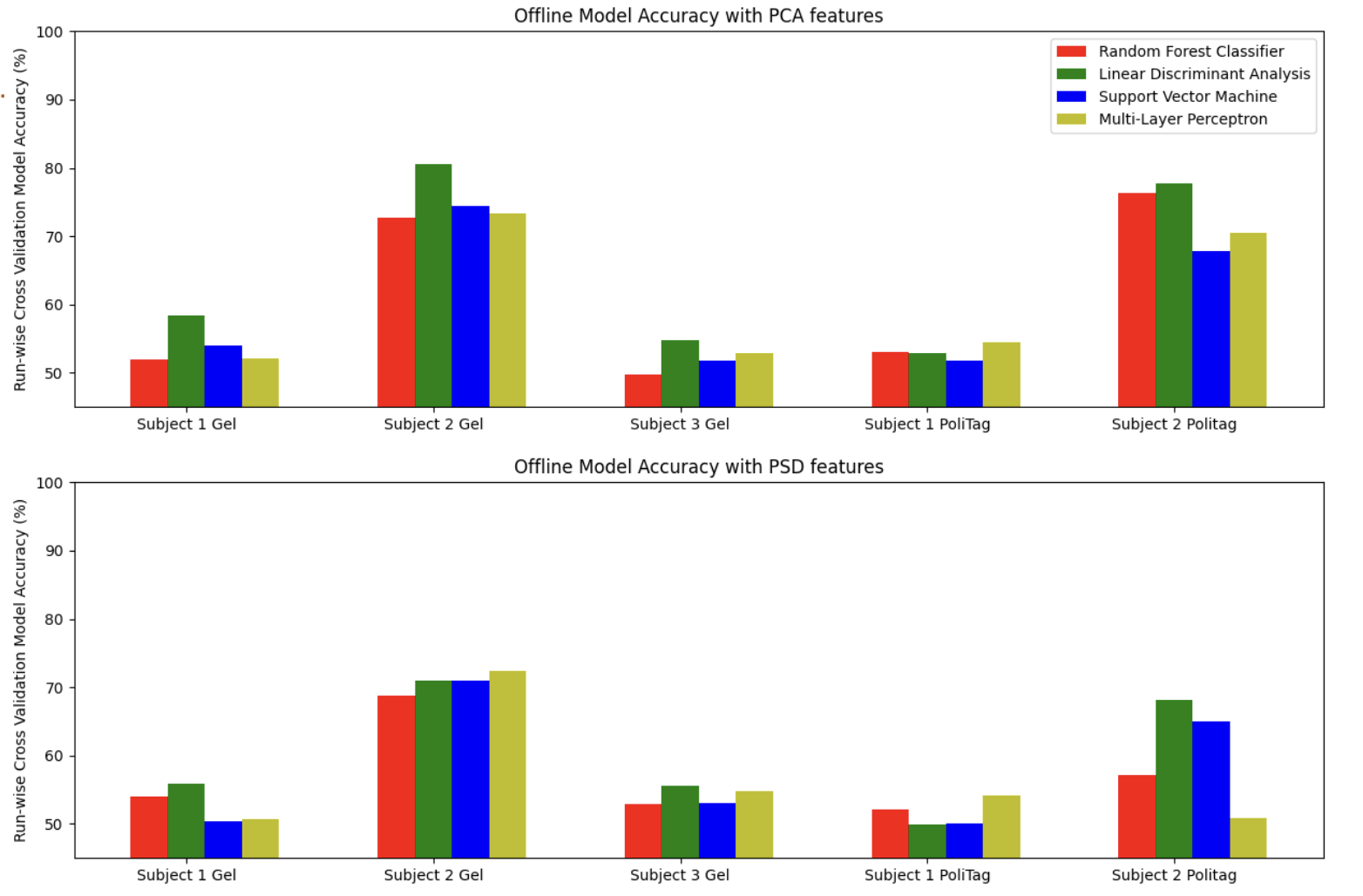}
\caption{Run-wise cross validation of four different machine learning models trained across each subject and sensor type. The top plot was generated using PCA to extract features and the bottom plot using PSD to extract features. In all cases, only the offline session was used.}
\label{fig_10}
\end{figure*}

Run-wise cross validation was performed using the 4 offline runs for each subject and for each sensor type. PCA or PSD features were extracted as described in section III.C.1 and III.C.2 and used as input. The results of this experiment can be seen in Figure \ref{fig_10}

Overall, the best performing model was clearly the Linear Discriminant Analysis model. The Multi-Layer Perceptron did out perform it at times, 
but these results overall seemed to indicate that the LDA model is the best choice. It is possible that with sufficient hyper-parameter tuning, models like Random Forest or Support Vector Machine could have achieved more competitive results, but tuning the hyper-parameters across multiple subjects and feature selection methods would have been prohibitively time intensive.

One thing worth noting from Figure \ref{fig_10} is that Subject 2's recorded data seems to be far more discriminable in general than that of Subject 1 or Subject 3. Regardless of the model, feature extraction method, or sensor used, Subjects 1 and 3 were hardly able to achieve results much better than random chance. Subject 2 on the other hand was able to reach as much as 80\% sample accuracy when the PCA features LDA model were used on the Gel recording.

Based on these results, we decided to use a Linear Discriminant Analysis model for training our online decoder. All subjects and sensor variations are to be training using this model, with hyper-parameter tuning.

\section{Online Evaluation}

With our feature extraction method and model selected, we build a decoder for each subject and sensor combination. After exploring the data from our subjects, we noticed that Subject 3's online gel sessions had not been properly recorded. We also noticed that Subject 1's online POLiTag sessions were missing runs. For the sake of consistency, those two are dropped and the remaining three, Subject 1 Gel, Subject 2 Gel, and Subject 2 POLiTag, are used for online evaluation. 

\subsection{Decoder Training}

To train an online decoder, we utilize all four runs from the offline session. 
The data from these runs is pre-processed as described in section III.A and then PCA features are extracted using the optimized number of PCA components as described in section III.B.1. 
Next a Linear Discriminant Analysis model with the Singular Value Decomposition optimizer is fit to the transformed offline run data. 
This fit is done at the sample level, where each sample is one of the 1s windows. This decoder is used in the sample-level evaluation and in the trial-level evaluation as described below.

\subsection{Sample-level results}

When the decoder was trained on the offline runs, we first performed a PCA transform of the data. The result of this PCA transform was not only the transformed data but transformer itself. In order for this experiment to work, we need the online runs to be transformed into the exact same feature space as the training data. Therefore, we keep the transformer that was used and apply it to the online runs corresponding to the offline run which generated it. These need to be kept track of between subject and sensor types because the optimal PCA value was different for each subject and sensor type. Subject 1 Gel's best PCA value was 520, Subject 2 Gel's best value was 760, and Subject 2 POLiTag's best value was 160.

To begin, the decoders were tested on online session 1 and online session 2. This testing happened at the sample level and only after the correct PCA transform had been applied to the test data. Following this, each decoder was trained again from scratch, however this time both the offline session and online session 1 was used as the training data to the decoder. This simulates the effect of fine-tuning the model on the online data collected. This fine-tuned decoder is then tested on online session 2 to see if the fine-tuning affects the performance. Additionally, a new PCA transform is computed based on the combined offline and online data, and this new transform applied to online session 2 before evaluating. 

\begin{figure}[!h]
\centering
\includegraphics[width=3.4in]{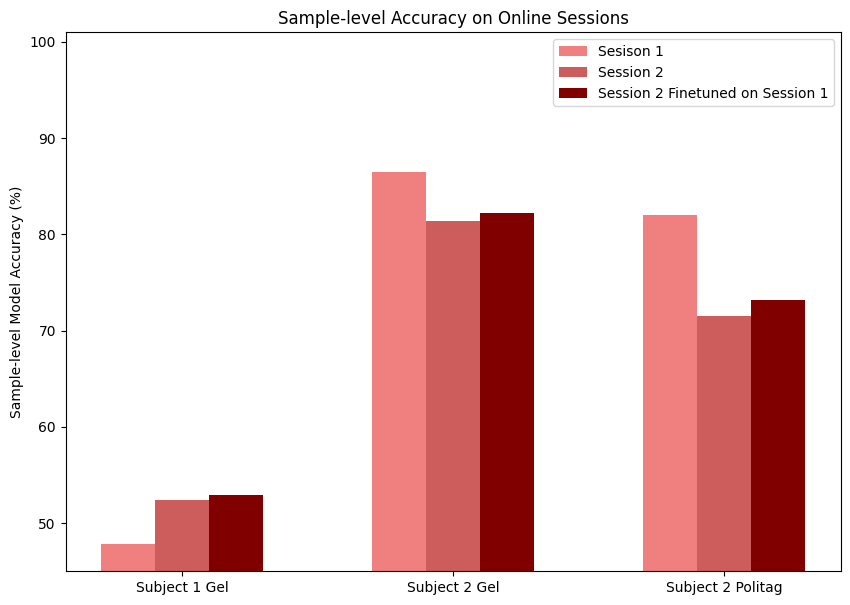}
\caption{Sample-level accuracy of the decoders on online sessions}
\label{fig_11}
\end{figure}

Figure \ref{fig_11} describes the results of this experiment. 
The performance was on the online sessions was comparable to that of the offline cross evaluation. In some cases the accuracy even increased. 
This is a good sign that we have learned representative features that generalize well across time, and that we have not over-fitted to the training data.

Additionally, our model fine-tuning seemed to have been successful. 
In all three cases, the performance of the decoder fine-tuned on online session 1 outperformed that of the decoder that was only trained on the offline data. 
This is unsurprising as the larger dataset likely allows the model to learn a more representative distribution of the features.

\subsection{Trial-level evaluation}

In addition to evaluating our models on the online data across just each sample, we also want to evaluate them across an entire trial. In the sample-level analysis, a session is made up of 60 trials which generate at most 63 windows which translates to 3780 samples which are evaluated independently. In the trial level analysis, we look only at the samples that fall within one trial and give one prediction for the entire trial.

To come up with this trial-level model prediction, we use an evidence accumulation framework. There are many ways to achieve this, but ours is as follows: First a threshold value is chosen. This can be anything between 0 and 1.0. Then a step value is chosen, typically something smaller than 0.1. We begin with an Evidence Value (EV) of 0.0 and begin by looking at the first window of a trial. The model gives it's prediction: Left or Right. If the model predicts Right, we add the step value of the Evidence Value. If the model predicts Left, we subtract the step value from the EV. With this setup, we will take a step left (negative) or a step right (positive) at each window/sample. We continue this for each window in the trial until the absolute value of the EV is greater than the threshold value, at which time we check the sign of the EV predicting right for the trial if the value is positive and predicting left for the trail is the value is negative. If the threshold is not surpassed by the time all windows in a trial are seen, the trial is reported as a timeout.

To discover the best parameters for step value and threshold value, we perform a grid search for each individual. We chose a range of values for each and used the percentage of correct, incorrect, and timeout predictions to chose a set of parameters that maximizes correct predictions while minimizing both incorrect and timeouts.

\begin{figure}[!h]
\centering
\includegraphics[width=3.4in]{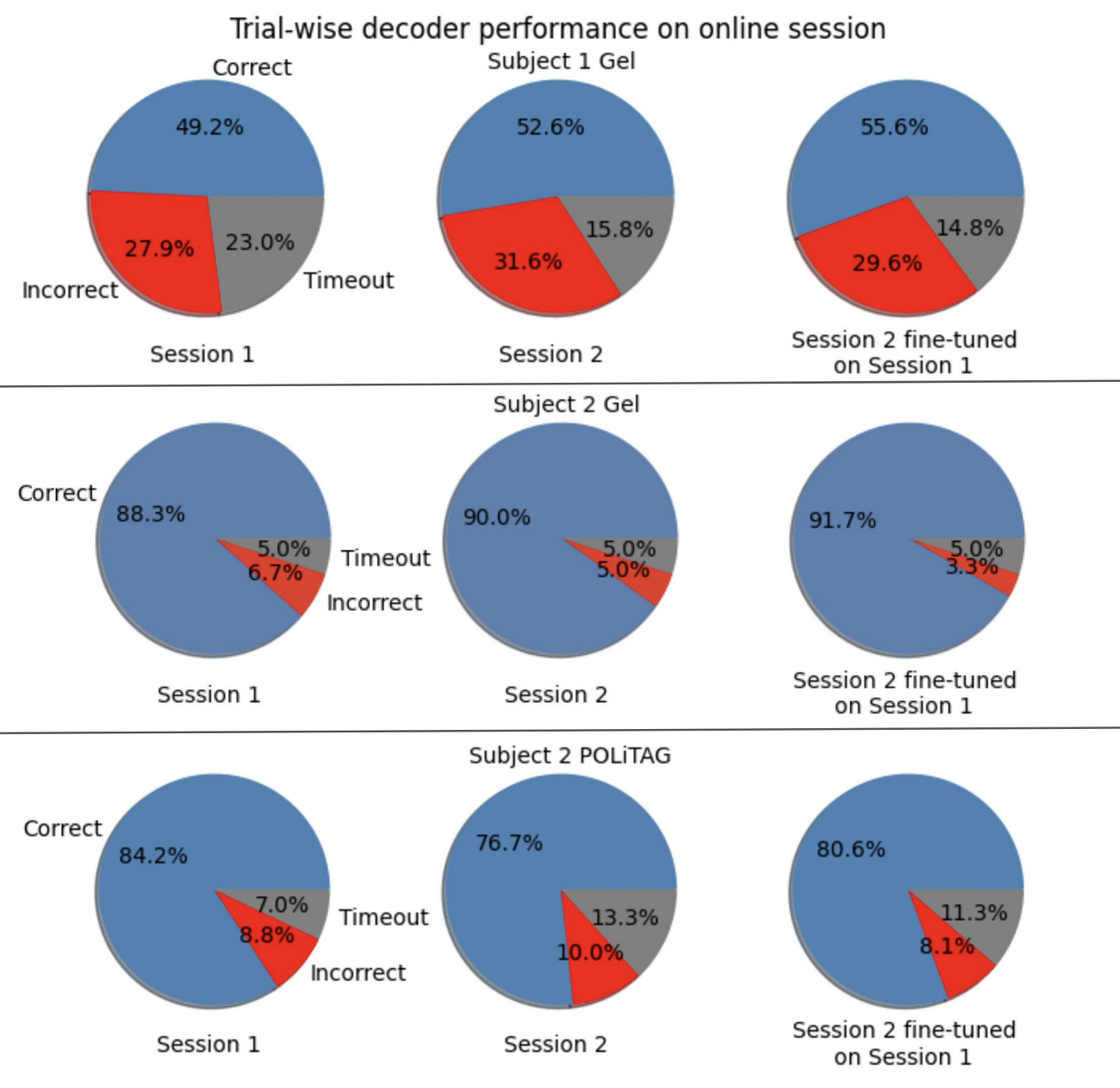}
\caption{Trial-level results of the decoders on online sessions}
\label{fig_12}
\end{figure}

The results of this experiment are shown in Figure 12. They seem to line up well with that of of the sample-wise performance. It can be seen that the performance of subject 2 was much better that subject 1. Furthermore, the results of session 2 improved when fine tuned on session 1 data.

\section{Conclusion}
In this motor imagery experiment, the objective was to build a decoder for a subject that could correctly predict the imagined direction as either left or right. There were several precautions that were taken care of right from the experimental setup to building the decoder to ensure that the data is not polluted. To eliminate unwanted noise, we made sure there was no such equipment near the BCI that could produce electromagnetic interference with the readings. During the experiment, we made sure that the subject did not face any distractions while focusing on the task. Once all the data for all the sessions was collected, it was properly filtered to improve its signal to noise ratio. Since there were too many features,a PCA tradeoff was also performed. As part of feature selection, a comparison with PSD was also done. Next, several ML models were experimented with. To begin with, a convolutional neural network model was built with varying depths, but it wasn't performing at par. Several scikit learn models were also compared and LDA seemed to give the best results. The decoder accuracies varied with subjects more that the kind of model that was selected. Since the training of models was based on subject specific data, so any error while recording the runs could lead to poor decoders for specific subjects. Finally, for online evaluation both sample level as well as trial level results were shown.
\section{Future Work}
As mentioned in a previous section, a different approach could be taken for training the convolutional neural network model. An image based approach could be used for both training and predicting. Generating a real time image plot of the EEG signal is not time or computation intensive, unlike training the CNN model that can be done after data collection.

\vfill

\end{document}